\documentstyle[11pt, aaspp4, psfig]{article}

\def\gta{\ifmmode {\mathbin{\lower 3pt\hbox   
    {$\,\rlap{\raise 5pt\hbox{$\char'076$}}\mathchar"7218\,$}}}
    \else {${\mathbin{\lower 3pt\hbox
    {$\rlap{\raise 5pt\hbox{$\char'076$}}\mathchar"7218\,$}}}
    $}\fi}
\def\lta{\ifmmode {\,\mathbin{\lower 3pt\hbox   
    {$\,\rlap{\raise 5pt\hbox{$\char'074$}}\mathchar"7218\,$}}}
    \else {${\mathbin{\lower 3pt\hbox
    {$\rlap{\raise 5pt\hbox{$\char'074$}}\mathchar"7218\,$}}}
    $}\fi}
\def\kms{{\rm km\ s^{-1}}}
\def\Mpc{{\rm Mpc}}
\def\g{{\rm g}}
\def\cm{{\rm cm}}
\def\msun{{\rm M_\odot}}

\begin{document}

\title{Suppression of Gravitational Structure Formation by Cosmological 
Accretion Heating}

\author{M. Coleman Miller and Eve C. Ostriker}
\affil{Department of Astronomy, University of Maryland\\
       College Park, MD  20742-2421\\
       miller@astro.umd.edu,ostriker@astro.umd.edu}

\begin{abstract}
  
  As increasingly precise information about the spectrum of the cosmic
  microwave background fluctuations is gathered with balloon and
  satellite experiments, interest has grown in foreground sources of
  opacity affecting these observations.  One potentially important
  source is electron scattering produced by a
  post-recombination luminosity source, which would significantly attenuate
  the higher harmonics in the spectrum.  If such an ionization source
  exists, then it would also heat the universe, hence increasing the
  Jeans mass and suppressing early gravitational structure formation.  
  Here we consider the effects of such heating.  We concentrate on one type
  of ionization source: luminosity generated by accretion onto
  primordial compact objects.  We show that if such objects generate
  enough luminosity to affect the CMB power spectrum, then they would
  produce enough heat to prevent the formation of 1$\sigma$ collapsed 
  objects until
  $z\sim 5$, significantly less than the redshift at which baryonic 
  collapse could otherwise occur.  Such processes would leave signatures
  detectable by upcoming instruments such as NGST, SIRTF, and SWIFT.

\end{abstract}


\section{Introduction}

The advent of a data-rich era in cosmology is expected to produce
dramatically enhanced understanding of the early universe, from the initial
spectrum of perturbations and their processing before radiation decoupling,
to the formation of the first galaxies and stars.  The high angular
resolution power spectra of the cosmic microwave background (CMB) gathered
by current and future satellite, balloon, and ground-based experiments are
expected to provide precise information about many cosmological parameters,
such as the total curvature of the universe and the baryonic mass fraction.
These power spectra are also expected to be informative about a number of
foreground sources, such as clusters of galaxies and other large-scale
structure, and about the epoch of ionization of the intergalactic medium
(e.g., Tegmark et al.~2000).

Early results from BOOMERanG and MAXIMA data suggested, surprisingly,
that the second acoustic peak is significantly weaker than expected
(de Bernardis et al. 2000; Hanany et al. 2000).  More recent results
from BOOMERanG (Netterfield et al. 2001), DASI (Pryke et al. 2001)
and possibly MAXIMA (Lee et al. 2001) are consistent with the standard
$\Lambda$CDM cosmology.  However, the systematic uncertainties in
current data are significant enough that it is still not possible to
tell with precision whether the higher harmonics are at the expected
amplitude; this will need to wait for instruments with absolute
calibration such as MAP and Planck (e.g., Tegmark et al.~2000).
Therefore, despite the encouraging agreement with expectations, it is
worthwhile to calculate the impact that other effects have at both
high and low redshifts.
  
The CMB spectrum can strongly constrain ionization at high redshifts,
but some ionizing processes that may have only subtle effects (but
nonetheless measurable with MAP and Planck) at $z\sim 1000$
can have important consequences at lower redshifts.  One such effect
within the standard model has been
suggested recently by Peebles, Seager, \& Hu (2000): if there exists
a radiation source in the universe around the time of decoupling
($z\sim 1000$), the resulting enhanced density of free electrons at
high redshift would
Thomson scatter the CMB and suppress the second and higher peaks, in
addition to moving the first acoustic peak to larger angular scales.
Such sources could have many additional cosmological effects.  In
particular, if scattering is posited to suppress the second acoustic
peak (Peebles et al. 2000) then the luminosity responsible for
increased ionization would also heat the universe.  As a result, the
Jeans mass at early epochs would be raised, which in turn could affect
the redshift of collapse of the first nonlinear structures.

One candidate for an early ionization source is accreting compact
objects that exist by a redshift $z\sim 1000$, e.g.,  black holes
formed during the quark-hadron phase transition.  The first detailed
investigation of the effects of such objects was performed by Carr
(1981).  He showed that the radiation generated by accretion of ambient
gas onto primordial black holes could have a significant effect on the
thermal and ionization history of the universe,  potentially even
preventing the universe from entering a neutral phase.  Given large-amplitude 
baryonic fluctuations, massive black holes could have formed soon after
recombination;  Gnedin \& Ostriker (1992) and Gnedin, Ostriker, \& Rees (1995)
considered the possible consequences for thermal, ionization, and 
nucleosynthesis histories of early injection of an AGN-like spectrum from
such accreting sources.

An update to Carr's 
analysis was performed by Miller (2000), who used the existing limits
on electron scattering optical depth between decoupling and the current
epoch to place an upper bound on the contribution of primordial compact
objects to dark matter.  This limit is based on the effect such
scattering would have on the angular power spectrum (limits based on
distortions of the energy spectrum are much weaker; see, e.g.,
Griffiths, Barbosa, \& Liddle 1999).  Miller found that the product
$\Omega_{\rm CO}\epsilon_{-1}(M/10\,M_\odot)<4\times 10^{-3}$ (in the
notation of the current paper, $\xi<4\times 10^{-3}$), where
$\Omega_{\rm CO}$ is the fraction of closure density in compact objects,
$\epsilon_{-1}$ is the accretion efficiency divided by 10\%, and $M$
is the typical mass of the compact object.  In this paper we show that
current data imply $\xi\lta 5\times 10^{-4}$, which corresponds to a limit
on optical depth that is three times stronger than used by Miller (2000)
based on earlier data.

Here we explore the effect of ionizing sources on the gravitational
structure formation history of the universe. We show that if accreting
primordial compact objects generate enough ionization to significantly
affect the second and third acoustic CMB peaks, then they raise the
temperature of the universe to $T\gta 500$~K when $z\lta 30$. As has been
discussed by a number of authors (e.g., Ostriker \& Gnedin 1996; Valageas
\& Silk 1999; Haiman, Abel, \& Rees 2000; Gnedin 2000), an increase in
temperature can dramatically alter the progress of structure formation.  We
show that in the present scenario, the Jeans instability threshold redshift
for an $n-\sigma$ density fluctuation is reduced by $\sim 5-7$, compared to
the threshold redshift when no extra heating sources are present.  For
example, $1-\sigma$ fluctuations can collapse only at $z\approx 5$.

We find that the parameter $\xi=\Omega_{\rm CO}\epsilon_{-1}(M/10\,M_\odot)$
governs both the distorting effect on the CMB power spectrum and the
formation of large-scale structure.  Because the two effects are therefore
linked, we do not need to assume anything about the accretion efficiency
$\epsilon$ for the compact objects; instead, we may simply say that if
accretion onto compact objects has a given effect on the CMB power spectrum,
it will have a related (and much stronger) effect on the redshift at which
nonlinear density peaks may form in baryons.  If the accretion efficiency
is low (e.g., if the flow is governed by an ADAF, CDAF, or 
wind-dominated solution, for which $\epsilon$ may be less than $10^{-4}$; 
see, e.g., Narayan, Mahadevan, \& Quataert 1998; Ball, Narayan, \& 
Quataert 2001; Blandford \& Begelman 1999), all effects are
proportionally reduced, but what we emphasize here is the relation between
possible subtle effects at high redshift and significant effects at lower
redshift.

In \S~2 we calculate the ionization fraction produced by an accreting
compact object beyond the immediate vicinity of the compact object (``HII
region'') in the surrounding ambient region which contains many
such objects.  In order to allow for nonlinear mass concentrations, we
generalize the analysis of Miller (2000) to include accretion in a
region with density differing from the average density of the
universe.  In \S~3 we compute the temperature of the HII and ambient
regions by balancing accretion heating with Compton cooling off the
CMB, also assessing the importance of bremsstrahlung, atomic cooling,
and molecular cooling.  In \S~4 we determine the heating implied if
Thomson scattering is responsible for reducing the second peak to
presently-observed levels.  We then evaluate the effect this level of
heating would have on the formation of self-gravitating structures.
In \S 5, we discuss the effects of more general ionization and heating,
and present our conclusions.

\section{Ionization Fraction}

In this section we focus on accreting primordial compact objects as
the source of early ionization and heating.  In \S~5 we will revisit
the heating effects of more general early ionization sources.  The
physical picture is of compact objects of typical mass $M$ capturing
matter via Bondi-Hoyle accretion from a surrounding medium.  For
generality we consider a medium with an ambient Hydrogen number
density $n_{\rm amb}$ and temperature $T_{\rm amb}$ that may differ
from the average density ${\bar n}$ and temperature ${\bar T}$ of the
universe.  This will allow us to treat accretion in underdense or
overdense mass concentrations.  The luminosity generated by accretion
will produce an HII region immediately around each compact object, in
which the ionization fraction $x\approx 1$.  Let the number density
and temperature inside the HII region be $n_{\rm HII}$ and $T_{\rm
  HII}$, respectively.  As discussed by many authors (e.g., Silk 1971;
Carr 1981; Miller 2000), the hardness of the spectrum produced by an
accreting compact object implies that its ionization effect is felt
over much greater distances than the radius of its Stromgren sphere,
and the resulting ionization fraction decreases much more slowly with
distance than is typical around an early-type star (as a power law,
not exponentially).

Miller (2000) calculated the average ionization fraction for a 
differential luminosity spectrum typical of accreting black holes
in our Galaxy,
$dL(E)/dE\propto E^{-1}\exp(-E/E_{\rm max})$ (the range of spectra observed
from stellar-mass black holes all give similar
answers for the ionization fraction; see Miller 2000 for a more detailed
discussion).  Typically
$E_{\rm max}\sim 100$~keV.  Over a region
large compared to the radius of the Stromgren sphere, the average
ionization fraction at redshifts $z\lta 100$ is
\begin{equation}\label{eq:xalg}
{\bar x}\approx 3\left[E_{\rm max}\over{3E_0\ln(E_{\rm max}/E_0)}\right]^{1/2}
\left(R_s\over{R_{\rm sep}}\right)^{3/2}\; .
\end{equation}
Here $E_0$=13.6~eV is the
ground-state ionization energy of hydrogen and the number of compact objects
per cosmological volume is $(4\pi R_{\rm sep}^3/3)^{-1}$.  
In terms of the fraction
$\Omega_{\rm co}$ of closure density in compact objects and the redshift $z$, 
\begin{equation}
R_{\rm sep}=8\times 10^{20} \,{\rm cm}(1+z)^{-1}
\left({M\over 10 M_\odot}\right)^{1/3}
\left({n_{\rm amb}\over {\bar n}}\right)^{-1/3}\Omega_{\rm co}^{-1/3}\; ,
\end{equation}
where we adopt $H_0=70 \kms\Mpc^{-1}$.  
The Stromgren radius is here defined as
\begin{equation}
R_s\equiv\left(2L\over{3\pi\alpha n_{\rm amb}^2E_{\rm max}}\right)^{1/3}\; .
\end{equation}
Here $\alpha=1.5\times 10^{-12}T_3^{-0.75}$~cm$^3$~s$^{-1}$ is the
recombination coefficient to the $n\geq 2$ states of hydrogen (Hummer 1994);
recombination to the $n=1$ state produces photons that are absorbed
almost immediately, taking the standard ``Case B" assumption.  We scale
the temperature as $T=10^3T_3$~K, because we find typical temperatures
$T\sim 10^3$~K at the redshifts of interest.  This form for the recombination
coefficient is accurate to better than 10\% for all $T>10^2$~K, and to
$\sim 1-2$\% for $10^3$~K$<T<10^5$~K (Hummer 1994).
The Bondi-Hoyle accretion rate is ${\dot M}=4\pi\lambda_s(GM)^2
\rho c_s^{-3}$, where for a $\gamma=5/3$ gas the accretion eigenvalue
is $\lambda_s=0.25$.  This produces a luminosity per source $L$ given by
\begin{equation}\label{eq:lum}
L\approx 3\times 10^{27} \,{\rm erg\ s}^{-1}\epsilon_{-1}
(1+z)^3\left(M\over{10 M_\odot}\right)^2 
\left({n_{\rm HII}\over{\bar n}}\right)T_{\rm HII,3}^{-3/2}\; ,
\end{equation}
adopting a cosmological baryon density $\bar \rho = 3.6\times 10^{-31} 
\g~\cm^{-3} (1+z)^3$ from Tytler et al (2000).
The accretion efficiency is defined as $L\equiv 0.1
\epsilon_{-1}{\dot M}c^2$.  The
Stromgren radius is then
\begin{equation}
R_s\approx 4.4\times 10^{19}\cm\, (1+z)^{-1}
\epsilon_{-1}^{1/3} \left({E_{\rm max}\over 10^4 E_0}\right)^{-1/3}
\left(M\over{10M_\odot}\right)^{2/3}\left(n_{\rm HII}\over{\bar n}
\right)^{1/3}\left(n_{\rm amb}\over{\bar n}\right)^{-2/3}
T_{\rm HII,3}^{-1/2}T_{\rm amb,3}^{1/4}\; .
\end{equation}
Pressure balance between the ambient medium and the HII region implies
$n_{\rm HII}T_{\rm HII}=n_{\rm amb}T_{\rm amb}$.  Collecting factors
and defining $\xi\equiv \epsilon_{-1}(M/10M_\odot)\Omega_{\rm co}$ as a
parameter quantifying the generation of luminosity by accretion, the average 
ionization fraction in the ambient medium for $z\lta 100$ is
\begin{equation}\label{eq:xnum}
{\bar x}\approx 0.7\, \xi^{1/2}\left[9\over{\ln(E_{\rm max}/E_0)}\right]^{1/2}
\left(T_{\rm amb}\over{T_{\rm HII}}\right)^{7/8}T_{\rm HII,3}^{-3/8}\; .
\end{equation}
The factor in brackets depends only on the spectrum, and is very close to
unity.  We note that the result from equations (\ref{eq:xalg}) and 
(\ref{eq:xnum}) is a factor  $[8/\ln(E_{\rm max}/E_0)]^{1/2}\sim 1$
times the result obtained by equating the total ionization at a mean rate
$n_{\rm co} L/(3 E_0)$ with recombinations at a mean rate $\alpha 
n_{\rm amb}^2 x^2$.
Because $L\propto n$ for energy generated via cosmic accretion, the 
resulting mean cosmic ionization is density independent except indirectly 
through the temperature.

\section{Balance of Heating and Cooling}

The formation of baryonic structure depends in part on whether the
nonlinear mass at a given redshift exceeds the Jeans mass at that
epoch.  In \S~4 we will derive the nonlinear mass.  The Jeans mass
depends on the typical temperature of the regions that
may undergo collapse.  We therefore need to compute this temperature in the
presence of accretion heating and ionization.

The temperature $T_{\rm amb}$ of the ambient medium is determined by
the balance of heating by accretion with cooling by a variety of processes.
Since ionization is not perfectly efficient (i.e., only
a fraction of available energy goes into ionization),  heating by accretion
is inevitable.  Cooling depends on a number of processes; candidates
include Compton cooling, bremsstrahlung, atomic recombination,
and molecular cooling.  In addition to considering ionization and thermal
balance in the ambient medium, we must also evaluate the latter in the HII
region. This is important because, at least for heating by
accretion, the luminosity per compact object depends on both the temperature
and the density inside the HII region (the Bondi radius is smaller than
$R_s$ for $M\lta 10^4 \msun$).  We first compute the HII region and ambient
temperatures $T_{\rm HII}$, $T_{\rm amb}$ by balancing accretion 
heating and Compton cooling in each region separately, assuming pressure 
balance applies at the interface.  We then demonstrate self-consistency by 
showing that if ionization is dominated by
radiative ionization (as opposed to collisional ionization) then cooling
by bremsstrahlung, atomic recombination, and formation of H$_2$ can
be ignored.  Only Compton cooling is effective.

{\it Balance of accretion heating and Compton cooling.} --- 
In general, the heating rate is related to the
ionization rate, with the heating rate per unit volume
$\Gamma = {\cal E}_{\rm T} \zeta (1-x) n_H$ for 
${\cal E}_{\rm T}$ the typical excess (thermal) energy in each ionization, and 
$\zeta$ the 
ionization coefficient.   As discussed by Carr (1981), through ionization
balance the heating rate may be related to the recombination rate
as $\Gamma={\cal E}_{\rm T} \alpha x n_H n_e$.  
Numerically, the heating rate at $z\lta 100$ is thus
\begin{equation}\label{eq:heating}
\Gamma\approx 6.2 \times 10^{-23} {\rm erg~cm}^{-3}~{\rm s}^{-1} 
x n_H n_e \left({{\cal E}_{\rm T}\over 2 E_0}\right) T_3^{-0.75}\; ,
\end{equation}
where the number densities here (and everywhere else in this paper) are
measured in cm$^{-3}$.
Carr (1981) estimates that there is $\approx 7$~eV of heating
per ionization within the HII region, in which primarily soft photons
interact due to their higher absorption cross sections.  In the ambient
medium far from the source, the soft photons have already been absorbed
and hence the photons are harder.
The number of 
ionizations per primary of energy $E$ is then
$\sim E/3 E_0$ for typical photon energies $\sim$1~keV or higher
(Dalgarno, Yan, \& Liu 1999), so that 
${\cal E}_{\rm T}\sim  E (E/3E_0)^{-1} - E_0 = 2 E_0$.

After decoupling but prior to structure formation the entire universe is
nearly transparent to CMB photons.  Scattering of these photons off of free
electrons provides a volume cooling rate
$4 \sigma_T a T_{\rm CMB}^4 k (T-T_{\rm CMB}) n_e/(m_e c)$, or 
\begin{equation}\label{eq:cmbcooling}
\Lambda_{\rm CMB}=5.6 \times 10^{-33} \ {\rm erg~cm}^{-3}{\rm \ s}^{-1} 
n_e (1+z)^4(T_{\rm 3}-T_{\rm CMB,3}) \; .
\end{equation}
At the redshifts $z\ll 1000$ relevant to our heating calculation,
the temperature in the HII region or ambient medium is much greater than
the temperature of the CMB and hence the factor in parentheses is
approximately $T_{\rm 3}$.  Inside the HII region, we set 
$n_H$ and $n_e$ to $n_{\rm HII}$, $x\approx 1$, and $T$ to 
$T_{\rm HII}$;  in the ambient region we set $n_H$ to $n_{\rm amb}$, $n_e$ 
to $x n_{\rm amb}$, and $T$ to $T_{\rm amb}$, and evaluate $x$ from 
equation (\ref{eq:xnum}).  From pressure balance with
the surrounding ambient medium, $n_{\rm HII}T_{\rm HII}=
n_{\rm amb}T_{\rm amb}$.

Equating the accretion heating and CMB Compton cooling rates 
(eqs. \ref{eq:heating},\ref{eq:cmbcooling}) in the 
HII region, we find
\begin{equation}\label{eq:THII}
T_{\rm HII,3} =  10 (1+z)^{-4/11} 
T_{\rm amb,3}^{4/11}
\left({n_{\rm amb}\over {\bar n}}\right)^{4/11}
\left({ {\cal E}_{\rm T, HII}\over 2 E_0}\right)^{4/11}\; .
\end{equation}
Similarly, thermal balance in the ambient region yields 
\begin{equation}
T_{\rm amb,3} =  70 (1+z)^{-4/7} \bar x^{4/7}
\left({n_{\rm amb}\over {\bar n}}\right)^{4/7}
\left({ {\cal E}_{\rm T, amb}\over 2 E_0}\right)^{4/7} \; .
\end{equation}
Combining this with 
our expressions from equation (\ref{eq:xnum}) for the ionization
fraction $\bar x$ 
in the ambient medium, and equation (\ref{eq:THII}) for the 
temperature of the the HII region, we solve to obtain 
\begin{equation}\label{eq:Tambsol}
T_{\rm amb,3} = 28  (1+z)^{-2/5}  \xi^{5/13}
\left({n_{\rm amb}\over {\bar n}}\right)^{2/5}
\left({ {\cal E}_{\rm T}\over 2 E_0}\right)^{2/5} 
\left({9\over \ln(E_{\rm max}/E_0)}\right)^{5/13}
\; ,
\end{equation}

\begin{equation}\label{eq:THIIsol}
T_{\rm HII,3} = 33  (1+z)^{-1/2}  \xi^{1/7}
\left({n_{\rm amb}\over {\bar n}}\right)^{1/2}
\left({ {\cal E}_{\rm T}\over 2 E_0}\right)^{1/2} 
\left({9\over \ln(E_{\rm max}/E_0)}\right)^{1/7} 
\; ,
\end{equation}
and 
\begin{equation}\label{eq:xsol}
\bar x = 0.19  (1+z)^{3/11}  \xi^{2/3}
\left({n_{\rm amb}\over {\bar n}}\right)^{-3/11}
\left({ {\cal E}_{\rm T}\over 2 E_0}\right)^{-3/11} 
\left({9\over \ln(E_{\rm max}/E_0)}\right)^{2/3} 
\; .
\end{equation}
The exponents are not exact, but are close approximations to the 
formally derived exponents.
We neglect possible differences between ${\cal E}_{\rm T}$ in the HII and
ambient regions.  Equations (11)--(13) apply at redshifts where
$T\gg T_{\rm CMB}$; as $z$ approaches decoupling, the effects of
accretion become negligible.  Figure~1 shows the high-redshift ambient
ionization boost for
$\xi=5\times 10^{-4}$ (which we find in \S~4.1 gives a good fit
to the second-year BOOMERanG and DASI data), compared to the standard model
in which there is no extra ionization and hence $\xi=0$.

\begin{figure}
\psfig{file=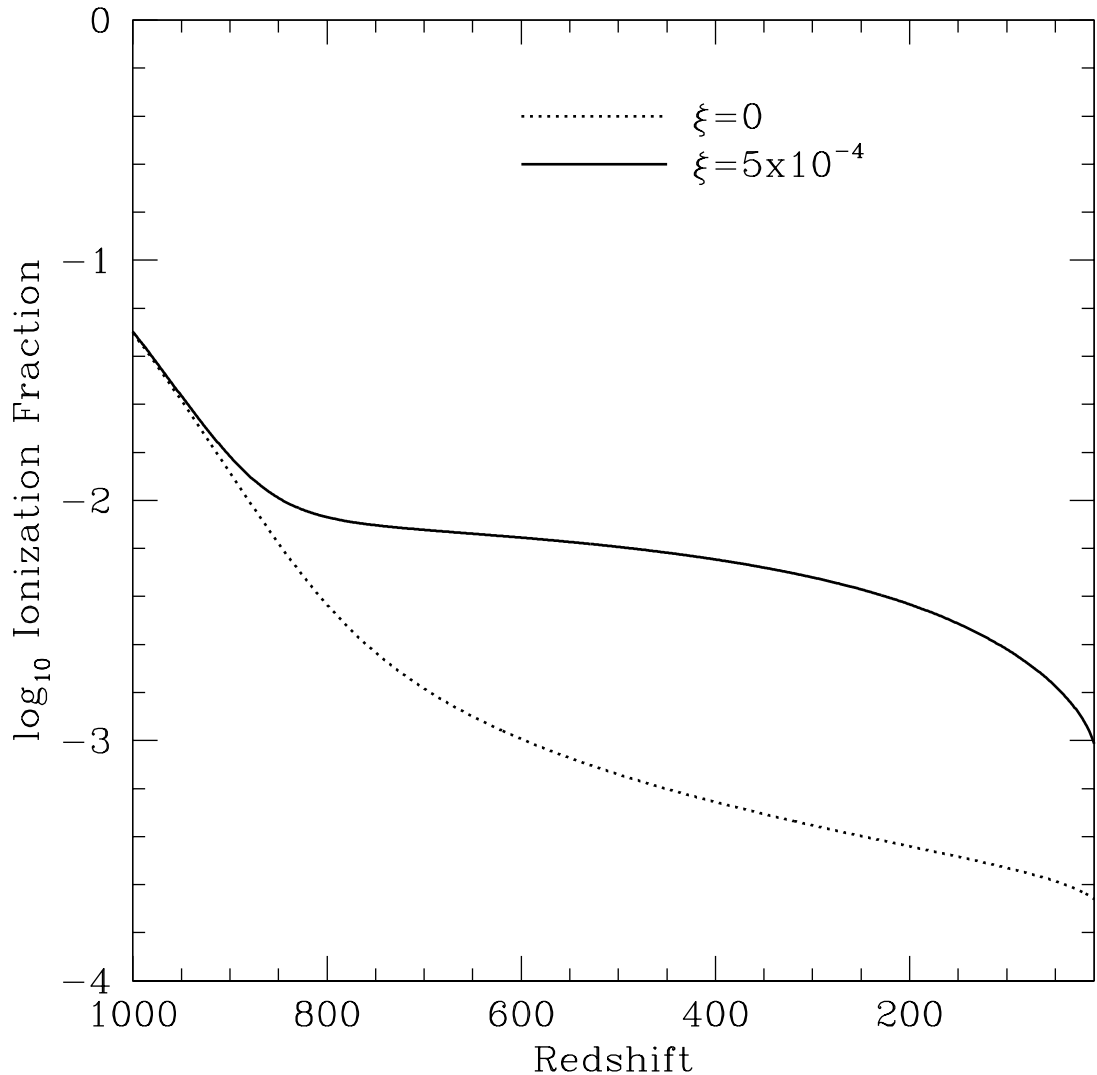,height=6.0truein,width=6.0truein}
\caption{Fractional ionization $x$ as a function of redshift, for
accretion by primordial compact objects ($\xi=5\times 10^{-4}$,
the best fit to the BOOMERanG and DASI data; see \S~4.1), and for no extra
ionization source ($\xi=0$).  Other cosmological parameters are chosen
as in \S~4.1.  The relatively small difference at high
redshifts implies that effects on the CMB power spectrum are small
(albeit measurable), but the large difference at lower redshift is
one consequence of the significant heating effect such sources can
have during the epoch of nonlinear structure formation.}
\end{figure}

We now consider three other cooling mechanisms: atomic recombination,
bremsstrahlung, and molecular cooling by formation of H$_2$.  For
each of these we assume that the ionization fraction $x$ is determined
by the radiative ionization balance calculated earlier.  In very
dense or hot media, collisional ionization will raise the ionization
fraction and therefore increase the effectiveness of these cooling
mechanisms.

{\it Atomic recombination}.---In solving for the ionization fraction
by equating the recombination rate with the radiative ionization rate
we have already effectively taken atomic recombination cooling into
account.  This is because the thermal energy, $kT<1$~eV, is much less
than the ionization energy of 13.6~eV.  Therefore, recombination plays
a minor role in cooling the medium.  If the temperature were high enough
that collisional ionization were important, then recombination might play
a major role, but here it can be neglected.

{\it Bremsstrahlung}.---The bremsstrahlung cooling rate is
\begin{equation}
\Lambda_{\rm bremss}=4.4 \times 10^{-26} {\rm erg\ cm}^{-3}~{\rm s}^{-1} 
xn_{\rm H} n_e T_{\rm 3}^{1/2}\; .
\end{equation}
The ratio of 
bremsstrahlung cooling to the heating rate is thus 
$\approx 6\times 10^{-4}\, T_{\rm 3}^{5/4}$, which is small for the 
present situation.

{\it Molecular cooling}.---No metals are present in the early universe,
so the only ways to form H$_2$ (by far the dominant molecule) are the
two-stage processes (1)~$H+e^-\rightarrow H^-+\gamma$ followed by
$H^-+H\rightarrow H_2+e^-$, and (2)~$H^++H\rightarrow H_2^++\gamma$
followed by $H_2^++H\rightarrow H_2+H^+$.  Puy et al. (1993) give
rates for these reactions, and (1) dominates by at least two orders
of magnitude over the temperature range of interest.  Moreover, the
second step of process (1) is effectively instantaneous compared to
the first, so the reaction rate for $H+e^-\rightarrow H^-+\gamma$
determines the overall rate of H$_2$ formation.  This rate is
$\alpha_{\rm mol}=10^{-15} ~{\rm s}^{-1}~{\rm cm}^3 T_{\rm 3}$.
Assuming as an upper bound that all $\sim$10~eV released in this
process escapes to infinity, the cooling rate in the (low-ionization) 
ambient medium is 
\begin{equation}
\Lambda_{\rm mol}=10~{\rm eV}\times \alpha_{\rm mol} n_e n_{\rm amb}.
\end{equation}
The ratio of this molecular cooling to the mean accretion heating
is $2.7\times 10^{-4}\, x^{-1} T_3^{7/4} $.  From equations (\ref{eq:Tambsol}) 
and (\ref{eq:xsol}), this ratio becomes 
\begin{equation}
\Lambda_{mol}/\Gamma = 0.5  (1+z)^{-1}  
\left({n_{\rm amb}\over {\bar n}}\right)
\left({ {\cal E}_{\rm T, amb}\over 2 E_0}\right) 
\; .
\end{equation}
This is generally less than unity for the redshift range of interest, for 
regions that are not strongly overdense compared to the universe as a 
whole.  Once a region becomes significantly condensed 
($n_{\rm amb}/\bar n\gg 1$), molecular cooling begin to dominate other 
cooling terms (Lepp \& Shull 1984).  For the purposes of assessing the
{\it onset } of collapse, however, we will simply use the result
expressed in equation (\ref{eq:Tambsol}) obtained in the approximation 
that accretion heating and CMB Compton cooling are in balance.

\section{Effects on the Formation of Structure}

With the ambient temperature $T_{\rm amb}$ that we
derived in the previous section, we can calculate the Jeans mass
within a region of ambient density $n_{\rm amb}$ 
if we know the accretion parameter
$\xi=\epsilon_{-1}(M/10\,M_\odot)\Omega_{\rm co}$.  We can then
compare the Jeans mass as a function of redshift to the 
nonlinear mass scales at that redshift and hence evaluate, as a function
of $\xi$, whether baryonic condensations within 
nonlinear dark matter peaks can collapse.  In this section we
first compute the maximum value of $\xi$ compatible with current
CMB data, then derive
the nonlinear masses and determine the epoch of structure formation.

\subsection{Relation to the CMB power spectrum}

We use the code CMBFAST (version 4.0; Seljak \& Zaldarriaga 1996 and
subsequent papers) to generate a power spectrum and compare it to
the best current signal to noise data, the second-year BOOMERanG
(Netterfield et al. 2001) and DASI (Pryke et al. 2001) data.  
As an illustrative fit, we fixed all of the
cosmological parameters except for $\xi$ and then fit for $\xi$.
We assume $\Omega_b=0.05$, $\Omega_{\rm CDM}=0.29$, $\Omega_\Lambda=
0.66$ (and hence assume a flat universe), $\Omega_\nu=0$, 
$H_0=66$~km~s$^{-1}$~Mpc$^{-1}$, $T_{\rm CMB}(z=0)$=2.726~K, 
$Y_{\rm He}=0.246$, three generations of massless neutrinos and none
of massive neutrinos.  These numbers are consistent with fits of
recent CMB data that 
include priors based on large scale structure,
Cepheid measurements of the Hubble constant, and Type~Ia supernovae.
This fit is also consistent with big bang nucleosynthesis.
We also assume an $n=1$ initial perturbation
spectrum and adiabatic initial conditions, and do not include tensor
perturbations.  We included the additional ionization by modifying the
routine ``recfast.f''.  At high redshift, the temperature of the HII region
is no longer much greater than the temperature of the CMB, and hence
for high $z$ we solve for the ionization numerically without using this 
assumption.  

We plot the resulting curves for no accretion ($\xi=0$), $\xi=5\times
10^{-4}$, and $\xi=10^{-3}$ against the second-year BOOMERanG and DASI data
in Figure~2.  The vertical bars are the estimated
statistical uncertainties in measurement.   Netterfield et al. (2001)
estimate that there is a 20\% systematic uncertainty in the points, and
Pryke et al.~(2001) estimate that there is a 16\% uncertainty in the
power measurements, but for
this figure we have plotted the raw output from CMBFAST against the data,
including only statistical
errors.  Note that there are also substantial systematic errors contributed
by beam uncertainties, particularly for $\ell>400$ (e.g., Netterfield et al.
2001), which have not been included in this figure.  Clearly, the power
spectrum is fairly sensitive to $\xi$ in this range.  To get a rough idea
of the goodness of fit, we made a simplistic comparison of the model with
the data at the center of each $\ell$ range, allowing an overall shift up
to $\pm$20\% but not varying the data to accommodate beam uncertainties.
The resulting $\chi^2$ values compared against the BOOMERanG data
are: 39.1 for $\xi=0$; 26.2 for $\xi=5\times
10^{-4}$; and 65.7 for $\xi=10^{-3}$.  In each case there are 17 degrees of
freedom, since there are 19 data points and $\xi$ and the overall scale are
varied. The best value of $\xi=5\times 10^{-4}$ gives a fit acceptable at
the 7\% level, not including the beam errors, which make all of the curves
significantly more compatible with the data. We note that the best-fit
value of  $\xi \equiv \epsilon_{-1} \Omega_{\rm co} (M/10\,M_\odot)$ is
compatible  with a low  value of the accretion efficiency $\epsilon$ and/or
a low value of the  fraction of closure mass in compact objects
$\Omega_{\rm co}$, provided the compact objects are relatively massive.

\begin{figure}
\psfig{file=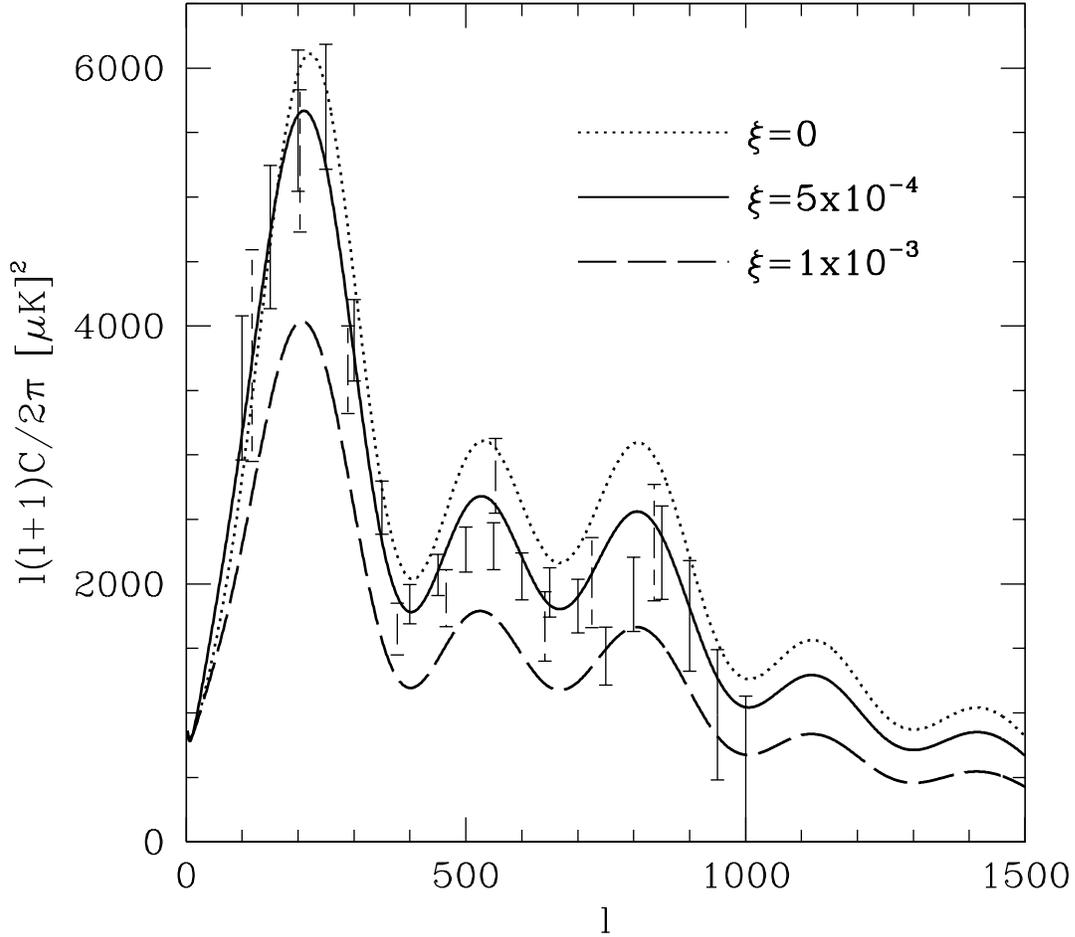,height=6.0truein,width=6.0truein}
\caption{Comparison of CMBFAST output, using the compact object accretion
model, to the second-year BOOMERanG data (Netterfield et al. 2001, solid error
bars) and DASI data (Pryke et al. 2001, dashed error bars),
including statistical but not systematic errors.   Here only $\xi$ is
varied, and other cosmological parameters such as $H_0$ and
$\Omega_\Lambda$ are fixed at the values indicated in the text.  When a
$\pm 20$\% shift is allowed, $\xi=0$ and $\xi=5\times 10^{-4}$ fit well,
but $\xi=10^{-3}$ still gives too low a first peak. Note that the amplitude
of the second peak is quite sensitive to $\xi$.  The best value,
$\xi=5\times 10^{-4}$, gives an acceptable fit: for example, compared to
the BOOMERanG data, $\chi^2$=26.2 for 17
degrees of freedom, including an overall systematic shift of 5\% but not
including beam errors.}
\end{figure}

\subsection{Growth of density perturbations}

When density perturbations are linear, the relative amplitude $D(z)$ of a
density perturbation $\delta = (\rho-\bar\rho)/\bar\rho$ evolves with the
overall scale factor $a \equiv (1 +z )^{-1}$ of the universe according to 
\begin{equation}
D(z)=E(z){5\Omega_m\over 2}\int_z^\infty {1+z^\prime\over{
E(z^\prime)^3}}dz^\prime \; 
\end{equation}
(Heath 1977).
Here $E(z)=H(t)/H_0=\left[\Omega_m(1+z)^3+\Omega_r(1+z)^4+\Omega_\Lambda +
  \Omega_k(1+z)^2 \right]^{1/2}$, and $\Omega_m$, $\Omega_r$,
$\Omega_\Lambda$ denote the present contributions to $\Omega$ from
matter, radiation, and the cosmological constant, and $\Omega_k = 1-
(\Omega_m + \Omega_r + \Omega_\Lambda)$ is the overall curvature
(our naming convention corresponds
to that of, e.g., Peebles 1993).  
For the regime of interest, we shall neglect the $\Omega_r$ term.  
The power spectrum of small amplitude fluctuations at redshift $z$ thus
evolves under linear growth according to
\begin{equation}
P(k,z)=\left[D(z)\over{D(0)}\right]^2P_0(k)\; ,
\end{equation}
where $k$ is the comoving wavenumber of a fluctuation.  

To obtain the
power spectrum shape and normalization, we follow the development of 
Bunn \& White (1997) and Hu \& Sugiyama (1996).  We
assume that the power law slope of the initial perturbations is
$n=1$, that the universe is flat $\Omega_k=0$, and that $k$, the comoving
wavenumber, is measured in Mpc$^{-1}$.  Bunn \& White (1997) write
the present-day power spectrum as $\Delta_0^2(k)=k^3P_0(k)/2\pi^2=\delta_H^2
\left(k\over{H_0/c}\right)^{3+n}T^2(k)$, with 
$\delta_H\approx 2\times 10^{-5}\Omega_m^{-0.73}$.  The transfer
function is given by (Bardeen et al. 1986)
\begin{equation}
T(q)=T\left(k\over{h\Gamma}\right)={\ln(1+2.34q)\over{2.34q}}
\left[1+3.89q+(16.1q)^2+(5.46q)^3+(6.71q)^4\right]^{-1/4}
\end{equation}
where $h\Gamma\approx\Omega_mh^2$.  From Carroll, Press, \& 
Turner (1992), a good approximation to the present growth factor is
\begin{equation}
D(0)\approx {5\Omega_m\over{2\left[\Omega_m^{4/7}-\Omega_\Lambda
+(1+\Omega_m/2)(1+\Omega_\Lambda/70)\right]}}\; ,
\end{equation}
so for $\Omega_m=0.3$ and $\Omega_\Lambda=0.7$, $D(0)\approx 0.8$.

The potential for collapsed objects of a given mass to form depends 
first of all on whether perturbations of the corresponding physical 
scale have grown sufficiently to become nonlinear.  The variance
in the fluctuation spectrum at a given mass $M$ (or comoving length
scale $R$) is related to the power spectrum at redshift $z$ 
by the following equation (e.g.  Barkana \& Loeb 2000):
\begin{equation}\label{eq:sigma}
\sigma^2(M)=\sigma^2(R)=\int_0^\infty {dk\over{2\pi^2}}k^2P(k,z)
\left[3j_1(kR)\over{kR}\right]^2\; ;
\end{equation}
we adopt a top-hat filter $j_1(x)=(\sin x-x\cos x)/x^2$.

A top-hat perturbation on comoving scale $R$ collapses in an
Einstein-de Sitter universe at the point when its overdensity as
predicted by {\it linear} theory reaches a value $\delta_c=1.686$
(e.g. Peebles 1993); this critical overdensity has only a weak
dependence on cosmological parameters.  The typical redshift for a 
1-$\sigma$ peak at mass scale $M$ to collapse is therefore given by the
implicit solution of $\sigma(M)=\delta_c$ for $z$ using equation 
(\ref{eq:sigma}); the collapse redshift for a mass scale $M$ residing in
a 2-$\sigma$ peak is given by solution of $\sigma(M)=\delta_c/2$, and 
so on.

Whether the local baryons collapse into a given dark matter halo depends on 
the halo mass relative to the Jeans mass:  baryons collapse with the 
halo only if the halo's gravitational potential ($\sim G M^{2/3}\rho^{1/3}$)
exceeds the specific thermal energy of baryons ($\sim k T/\mu$).
If we assume spherical density
perturbations then
the Jeans mass for a cloud of temperature $T$ (and sound speed
$c_s$) and total matter density 
$\rho_{\rm tot}$ (including both baryons and cold dark matter) is
\begin{equation}\label{eq:mjeans}
M_J={1\over 6}\pi\rho_{\rm tot}\left(\pi c_s^2\over{
G\rho_{\rm tot}}\right)^{3/2}
\approx 7\times 10^5\, M_\odot (\rho_{\rm tot}/\rho_{\rm amb})^{-1/2}
T_{\rm amb,3}^{3/2}n_{\rm amb}^{-1/2}
\end{equation}
for a cosmological abundance of helium.
Here $n_{\rm amb}$ is measured in cm$^{-3}$ and $\rho_{\rm amb}$ is
the ambient density in baryons only.  If we assume that during
collapse the ratio of total mass to baryonic mass in a mass concentration
remains approximately constant at the average value (roughly 9),
then $(\rho_{\rm tot}/\rho_{\rm amb})^{-1/2}\approx 1/3$ and
$M_J\approx 2\times 10^5\, M_\odot T_{\rm amb,3}^{3/2}n_{\rm amb}^{-1/2}$.
In the standard model, where there is no extra source of
heating, the minimum halo mass for collapsed baryonic objects 
including shell crossing by dark matter and
other effects is (Barkana \& Loeb 2000)
\begin{equation}
M_{\rm min}=5\times 10^3\,M_\odot\left(1+z\over{10}\right)^{3/2}
\left(\Omega_mh^2\over{0.15}\right)^{-1/2}
\left(\Omega_bh^2\over{0.022}\right)^{-3/5}\; ,
\end{equation}
which is almost identical to the value $M_{\rm J,\ stand}$ 
that the Jeans mass of equation
(\ref{eq:mjeans}) would take on in the absence of cosmic heating
(see eq. 41 of Barkana \& Loeb 2000).
If instead heating by accretion {\it is} included, 
then from the results of the previous section we obtain
\begin{equation}\label{eq:mjmod}
M_{\rm J,mod} \approx 8\times 10^{10}\, M_\odot (1+z)^{-2}
\xi^{7/12}(n_{\rm amb}/{\bar n})^{1/10}\,
\approx 8\times 10^8\,M_\odot\xi^{7/12}[(1+z)/10]^{-2}\; .
\end{equation}
Note that this mass is virtually independent of $n_{\rm amb}$.
Since $\xi\sim 5\times 10^{-4}$ fits the BOOMERanG data,
this implies 
$M_{\rm J,mod} \approx 1.0\times 10^7\,M_\odot[(1+z)/10]^{-2}$.  This
mass is vastly greater than it would be sans heating, and hence has
a major effect on structure formation.

Figure~3 exhibits this effect.
Here we show, as a function of
redshift, the nonlinear mass $M_{\rm nl}$ for 1-$\sigma$, 2-$\sigma$,
and 3-$\sigma$ peaks, along with the standard Jeans mass and the
Jeans mass modified by accretion heating (with $\xi=5\times 10^{-4}$).
The nominal collapse redshift for a given rarity of peak is
significantly less than it would be without heating.

\begin{figure}
\psfig{file=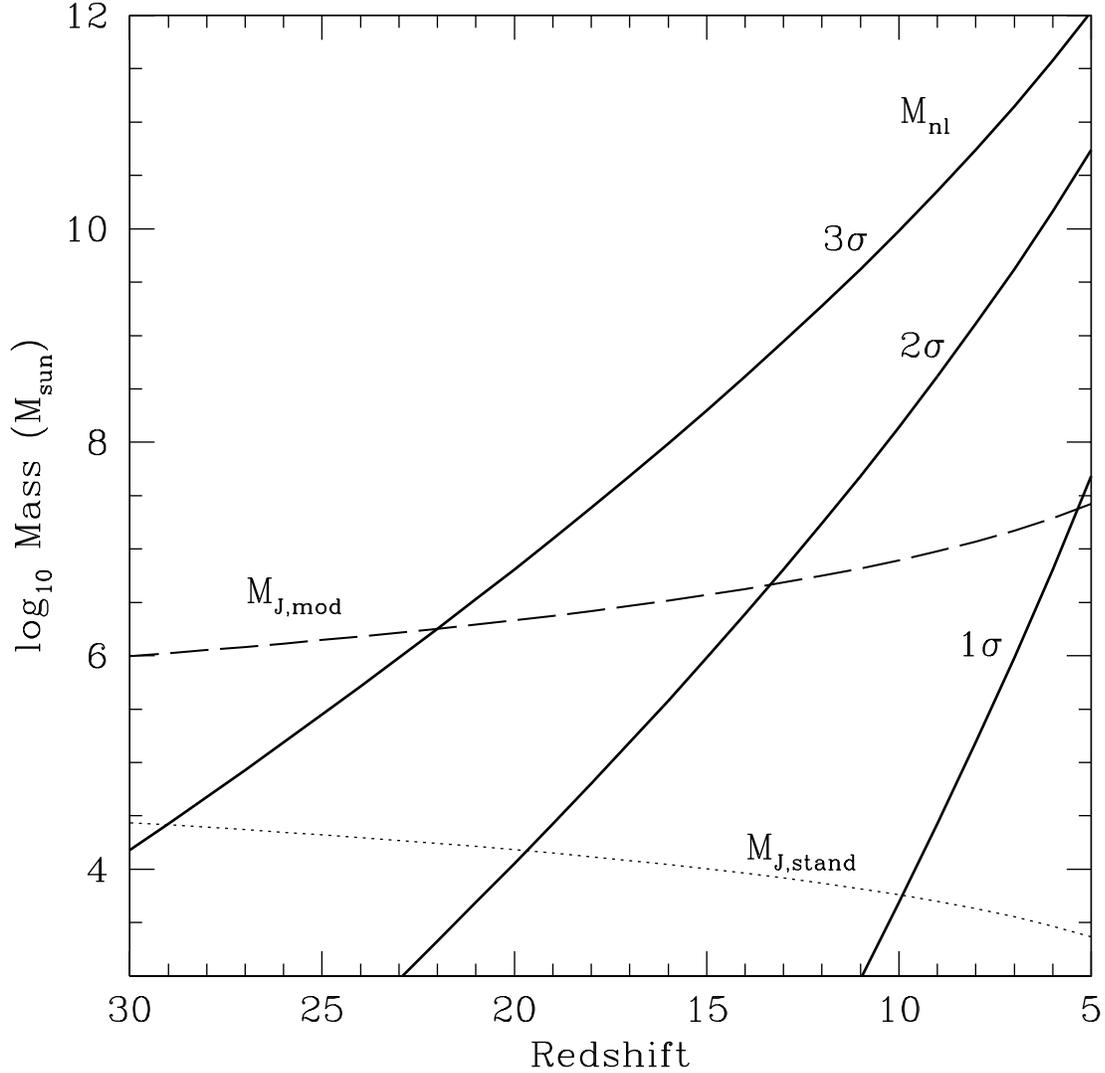,height=6.0truein,width=6.0truein}
\caption{Nonlinear mass and Jeans mass as a function of redshift,
with and without accretion heating.  The solid lines show the
nonlinear mass $M_{\rm nl}$ for 1$\sigma$, 2$\sigma$, and 3$\sigma$
peaks.  The dotted line ($M_{\rm J,stand}$) is the Jeans mass
assuming  no additional heating, and the long dashed line ($M_{\rm
J,mod}$) is the Jeans mass modified by heating ($\xi=5\times
10^{-4}$) compatible with the most recent BOOMERanG, MAXIMA, and DASI
data.  This figure shows that the effect of such
heating is to delay baryonic structure formation drastically; typically the
formation redshift of an n$\sigma$ peak is 5-7 less than what it would be
without the heating.}
\end{figure}

\section{Discussion and Conclusions}

We have shown that if accretion onto primordial compact objects
produces enough luminosity to increase significantly the ionization
fraction of the universe at $z\sim 1000$, then this luminosity has
a dominant effect on the thermal evolution of the universe at
lower redshifts.  In particular, the Jeans mass is raised by some three
orders of magnitude at $z\sim 10$, which would substantially delay baryonic
structure formation.  Even if the ionization produced by compact
objects is insufficient to affect the second CMB acoustic peak
strongly, the moderate dependence of the Jeans mass in equation
(\ref{eq:mjmod}) on the accretion luminosity parameter $\xi$ implies
that the presence of even a relatively small number of
(evenly-distributed) primordial compact objects could have a strong
effect on structure formation.

Effects on the observed cosmic background radiation other than on the
power spectrum are very subtle.  For example, the existence of extra
free electrons with a temperature elevated above the radiation temperature
will distort the photon energy spectrum.  To first order this can be
characterized by the Compton $y$ parameter, $y=\int d\tau\,k\Delta T/m_ec^2$.
However, even with the enhanced ionization and temperature, 
$y=2\times 10^{-9}$ when $\xi=5\times 10^{-4}$, compared with the 
limit $|y|<7\times 10^{-6}$ from
COBE measurements (Mather et al. 1999).  Thus, energy spectral distortions 
from scattering place far weaker limits on extra ionization than do limits
to changes in the power spectrum (see also Griffiths et al. 1999).
Another potential contributor to the energy spectrum is the luminosity
from the accretion itself.  However, this is also negligible.  From
equation~(\ref{eq:lum}), at $z\approx 10^3$ the luminosity per source
is roughly $10^{36}\,{\rm erg\,s}^{-1}\epsilon_{-1}(M/10\,M_\odot)^2$.
The total luminosity is $L\approx 2\times 10^{57}\,{\rm erg\,s}^{-1}
\epsilon_{-1}\Omega_{\rm CO}(M/10\,M_\odot)^2=2\times 10^{57}
\,{\rm erg\,s}^{-1}\xi(M/10\,M_\odot)$.  Taking redshifts into account
and assuming a radius of $\approx 10^{28}$~cm for the universe, the
observed flux would be $F\approx 2\times 10^{-6}\xi$~erg~cm$^{-2}$~s$^{-1}$.
Even for $\xi=1$ this is a factor of $\approx 10^3$ less than the flux
from a 2.7~K blackbody.  Taking into account that only a small fraction
of the accretion radiation will be near the blackbody frequencies and
that $\xi=5\times 10^{-4}$, the ratio of fluxes is $\approx 10^7$.  The
direct radiation is unobservable.  Thus, at high redshift the power spectrum
places the strongest limits on accretion by primordial compact objects.
However, as we have shown, at low redshifts the effects can be much more
dramatic.

An important difference between ionization by accreting primordial
objects and ionization by, e.g., the first generation of stars or
quasars is that the latter only exist after ordinary structure has already
formed, whereas the former ionize and heat the universe independent of
any structure formation.  Hence, accretion onto primordial objects, or
any other mechanism independent of structure formation at moderate 
redshift, can in principle delay any structure formation. Similar
effects from early stars or quasars are self-limiting, and thus may not 
prevent -- but nevertheless will regulate -- further growth of structure
(Ostriker \& Gnedin 1996).

More generally, any ionization source will also heat the universe.
For example, Peebles et al. (2000) analyzed a simple model in which a uniform
source of Ly~$\alpha$ photons exists early in the universe.  More hydrogen
atoms are therefore in the $n=2$ state, so they can be ionized more
easily.  The enhanced ionization produces more free electrons, which
scatter the CMB background, moving
the surface of last scattering to lower redshift and therefore moving
the first CMB peak to lower $\ell$ (larger angular scales) and decreasing
the amplitude of the second CMB peak.  In their model, they assume that
the rate of production of Ly~$\alpha$ photons per unit volume is
\begin{equation}\label{eq:peebles}
{dn_\alpha\over{dt}}=\epsilon_\alpha n_H H(t)\; .
\end{equation}
Here $n_H$ is the number density of hydrogen, $\epsilon_\alpha$ is a
free parameter, and $H(t)={\dot a}/a$ is the Hubble parameter at time $t$.
Peebles et al. (2000) found that with this model, $\epsilon_\alpha=10$
gave a good fit to the first-year BOOMERanG data. 

Let us assume that for each Ly~$\alpha$ photon produced by the sources
there is some heating as well, and that the energy that heats the 
universe is a fraction $f$ of the energy in the Ly~$\alpha$ photons
themselves.  At moderate and high redshifts with $\Omega_m=0.3$ and
$H_0=70$~km~s$^{-1}$~Mpc$^{-1}$, $H\approx 1.2 \times
10^{-18} \,{\rm s}^{-1} (1+z)^{3/2}$; the baryon density is 
$n_H = 1.5 \times 10^{-7} \cm^{-3} (1+z)^3$ from Tytler et al. (2000).
The energy of a Ly~$\alpha$ photon is approximately
10~eV, so the volume heating rate associated with equation (\ref{eq:peebles})
would be
\begin{equation}
\begin{array}{rl}
\Gamma&=f\epsilon_\alpha(10\,{\rm eV})n_HH(t)\\
&=3\times 10^{-36}\,{\rm erg\,cm}^{-3}{\rm s}^{-1}
f\epsilon_\alpha(1+z)^{9/2}\; .
\end{array}
\end{equation}
  From \S~3, CMB cooling dominates over recombination and molecular
cooling for the conditions of interest.  If the ionization fraction
is $x$ and the temperature of the ambient medium is $10^3T_{\rm amb,3}$~K,
the cooling rate is
\begin{equation}
\Lambda_{\rm CMB}=1\times 10^{-39}\,{\rm erg\,cm}^{-3}{\rm s}^{-1}
x(1+z)^7T_{\rm amb,3}\; .
\end{equation}
Equating the two, the temperature is $T_{\rm amb,3}=3000f\epsilon_\alpha
x^{-1}(1+z)^{-5/2}$.   If the hydrogen is completely
ionized ($x=1$), $\epsilon_\alpha=10$ from Peebles et al (2000)
implies a temperature of $1.5\times 10^4f$~K
at $z=20$; reducing the ionized fraction raises the temperature.  
Thus, unless the generation of Ly~$\alpha$ photons occurs
almost free of heating ($f\ll 0.01$), the heating present if this source
persisted at moderate redshift would raise the
temperature dramatically above what it would be in the standard model,
and hence increase the Jeans mass and delay structure formation.
The only type of ionization source that could significantly affect
the CMB power spectrum and yet not significantly affect structure
formation is one with a very steep redshift dependence.  If the
energy generation rate per volume is $\epsilon\propto (1+z)^n$
with $n>7$, heating effects at low redshift will be small.  Otherwise,
such heating will affect structure formation dramatically.

The richness of data available from upcoming cosmological experiments
will rapidly establish whether such ionizing and heating sources exist.
If, for example, the first nonlinear baryonic
objects did not form until $z\sim 15$ or more recently, evidence of this
should be readily apparent in data from many future satellites, including
NGST (see, e.g., Robinson \& Silk 2000) and SIRTF (e.g., Fazio,
Eisenhardt, \& Huang 1999), and in the redshifts and 
absorption line spectra from
distant gamma-ray bursts observed with HETE-2 and SWIFT (Wijers et al.
1998; Lamb \& Reichart 2000; Blain \& Natarajan 2000).  
In addition, as more data with better calibrations are
available from CMB experiments, the effects at higher $\ell$ will be more
sensitive to the presence of any extra ionization (Miller 2000).
In either case, the role of extra ionizing and heating sources in the
early universe will be clarified greatly in the next five years. 

\acknowledgements
We are grateful to N. Gnedin for useful information, and to 
J. Ostriker, D. Lamb, and A. Loeb for helpful comments on the manuscript.
We also thank the referee for suggestions that improved the clarity
of the paper.  This work was supported in part by NASA grant NAG 5-9756.

\end{document}